\documentclass[aps,prb,twocolumn,groupedaddress,showpacs,10pt]{revtex4-2}
\usepackage{graphicx}
\usepackage{color}
\usepackage{calc}
\usepackage{array}
\usepackage{amsmath, amssymb}
\usepackage{hyperref}

\newcommand{\unn}[1]{ \underline{\underline{#1}} }    
\newcommand{\vrd}{\vec}                               
\newcommand{\mxrd}[1]{\unn{#1}}                       

\begin{document}
\title{Ab initio Engineering of Topological Superconductivity in Substrate-Supported Magnetic Chains}
\author{Andr\'as L\'aszl\'offy$^{1,2}$}
\email{laszloffy.andras@itk.ppke.hu}
\author{Bal\'azs \'Ujfalussy$^{2}$}

\affiliation{$^1$Faculty of Information Technology and Bionics, Pázmány Péter Catholic University, H-1083 Práter u. 50/A, Budapest, Hungary}
\affiliation{$^2$HUN-REN Wigner Research Centre for Physics, Institute for Solid State Physics and Optics, H-1525 Budapest, Hungary}

\date{\today}

\begin{abstract}
In this paper, we employ a combination of several first-principles-based approaches to systematically scan chains of several types of magnetic atoms in different chain geometries on top of various overlayers on the Nb(110) surface, to develop a pathway to engineer topological Majorana edge states. Using atomistic spin-model calculations, we first identify a rich variety of magnetic configurations, including ferromagnetic, antiferromagnetic, spin-spiral, and exotic up-up-down-down states stabilized solely within a second-order Heisenberg model. In the next step, we solve the Bogoliubov--de Gennes equations using a band-theoretical method. Depending on the magnetic structure and atomic configuration, the superconducting spectra may exhibit topologically trivial minigaps, broad overlapping Shiba bands, or localized zero-energy edge states accompanied by sizable topological minigaps. In particular, Fe chains on Ag/Nb(110) are found to display pronounced signatures associated with topological superconductivity, including strongly localized zero-energy edge states and indications of Shiba band inversion. We further demonstrate that changes in the spin structure or a small amount of spatial relaxation can substantially modify the superconducting band structure and the localization properties of the edge states.

\end{abstract}

\maketitle

\section{Introduction}

Magnetic chains on superconducting substrates have attracted considerable interest as potential hosts of Majorana zero modes (MZMs), which may enable decoherence-free topological quantum computing. A magnetic impurity on a superconductor induces a Yu--Shiba--Rusinov (YSR) state \cite{yu1965bound,shiba1968classical,rusinov1969superconductivity}; in an atomic chain, these states hybridize into Shiba bands\cite{shiba1968classical} that can support topological superconductivity without external magnetic fields or fine-tuned gating. The resulting MZMs are protected by the particle-hole symmetry of the superconducting state and emerge at the ends of a topological superconducting phase through the bulk-edge correspondence\cite{nayak2008non,Alicea2012,Beenakker2020,Marra2022review,nyari_PRBI,Liang2011}.

Although MZMs have also been sought in two-dimensional systems, here we focus on finite chains of magnetic atoms on superconducting surfaces. Even in these comparatively simple heterostructures, however, MZMs have not been identified unambiguously. Scanning tunneling microscopy experiments have observed zero-bias peaks inside the superconducting gap in several candidate systems \cite{Jeon2017,Kim2018,Schneider2021a,Schneider2021b}, but such peaks alone do not constitute definitive evidence of MZMs.

A reliable theoretical description must therefore account for material composition, chain geometry, substrate coupling, spin-orbit coupling, and both structural and magnetic relaxation. First-principles methods are especially valuable because several of these quantities are difficult to determine in situ. In particular, Green's-function solutions of the Kohn--Sham--Dirac--Bogoliubov--de Gennes (KSDBdG) equations \cite{csire2018relativistic,Nyari2021} have reproduced key features of STM experiments \cite{Nyari2021,saunderson2022full,park2023effects,wu2023magnetic} and have shown that topological band inversion can be identified through quantities that are antisymmetric about the Fermi energy.

Two such quantities are the superconducting order parameter and the quasiparticle charge density of states (CDOS). The singlet order parameter changes its characteristic sign structure when a topological minigap is formed, thereby signaling Shiba-band inversion \cite{Ketterson1999,nyari_PRBI}. Similarly, defining
\begin{equation}
\textrm{CDOS}(\varepsilon)=e\left[\textrm{DOS}_e(\varepsilon)-\textrm{DOS}_h(\varepsilon)\right],
\end{equation}
particle-hole symmetry implies
\begin{equation}
\textrm{CDOS}(\varepsilon)=-\textrm{CDOS}(-\varepsilon).
\end{equation}
The spatially resolved or chain-integrated CDOS can therefore provide a complementary visualization of the topological character of the minigap \cite{nyari_PRBI}.

These diagnostics were previously applied to Fe chains in a continuum of spin-spiral configurations on Au-covered Nb(110) \cite{nyari_PRBI,laszloffy_PRBII}. Although a wide range of these configurations supported a topological band structure and MZMs, the actual ground state was found to be an out-of-plane ferromagnetic state without a minigap, both of which are consistent with experiment \cite{beck2023search}. This illustrates a central challenge: a topological state is technologically relevant only if it is compatible with a realizable, preferably ground-state, magnetic configuration.

A further requirement for topological quantum computation is the creation and manipulation of several MZMs, for example in T- or Y-shaped junctions. This raises the question of whether chains oriented along different crystallographic directions can simultaneously support suitable magnetic and topological states. On anisotropic surfaces such as Nb(110), direction-dependent atomic spacing modifies magnetic interactions and may therefore alter both the magnetic ground state and the formation of MZMs.

The present work addresses these issues in two steps. First, we search for combinations of magnetic-chain material and covering layer on Nb(110) that support an MZM in the magnetic ground state. Second, we examine whether such behavior can occur for different chain orientations. We therefore consider homogeneous Cr, Mn, Fe and Co chains in three different crystallographic orientations on Re, Ir, and Ag overlayers epitaxially grown on Nb(110).

To obtain material-specific and predictive results, we employ a fully first-principles workflow consisting of: (i) Relaxation of the spatial structure of both overlayer and chain atoms in a VASP calculation; (ii) self-consistent spin-polarized SKKR calculations in the relaxed geometry to obtain parameters for the next step; (iii) spin-dynamics simulations based on a classical spin Hamiltonian to determine the magnetic ground state; (iv) self-consistent SKKR calculations for that magnetic configuration; and (v) solution of the Dirac--Bogoliubov--de Gennes equations to obtain the electron- and hole-resolved local density of states and determine the topological properties and possible presence of MZMs \cite{csire2018relativistic}. This approach incorporates both geometrical and magnetic relaxation into the final superconducting-state calculation.

In the current paper, we performed calculations for a large variety of magnetic chains on heavy metal overlayers on the superconducting Nb(110) surface. The material of the overlayer was either Re, Ir, or Ag, which follows the 2D crystal structure of Nb in the calculations. Note that this assumption was verified experimentally in the case of a Ag overlayer\cite{ruckman1988}, 
a nearly stain-free epitaxial grown is expected for one monolayer of Re\cite{welander2010}, 
but for the Ir overlayer, a superlattice is formed where there are Ir atoms occupy only every second hollow sites.\cite{beck2022} The magnetic atoms, Cr, Mn, Fe, or Co are then placed in a closed-packed straight line (in hollow sites) along different crystallographic directions, $x$ ($\overline{1}10$), $y$ ($001$), or along the diagonal $u$ ($\overline{1}11$) direction. Note that the distance between neighboring magnetic atoms are $\sqrt{2} a_{\textrm{Nb}}$, $a_{\textrm{Nb}}$, and $\sqrt{3}/2 a_{\textrm{Nb}}$ in the case of the $x$, $y$, and $u$ crystallographic direction, respectively. The systems are then labeled as $X_{19}d~/~O\textrm{Nb}(110)$, where $X$ stands for the type of the magnetic atom, 19 is the number of magnetic antoms, $d$ displays the crystallographic direction, and $O$ is the material of the overlayer.

\section{Computational details}

As we mentioned in the introduction, the first step in our calculation is the optimization of the geometry, via a calculation with VASP. Each system was modeled as a $7 \times 7$ surface cell in a slab geometry consisting of three atomic layers of Nb, one atomic layer of the overlayer, and a magnetic adatom (in total 197 atoms). The chosen geometry ensures that the magnetic atoms does not interact with each other between the repetitive cells. In the (001) direction, a vacuum region with at least 10\AA~thickness was considered to avoid interaction between repetitive cells. We used the generalized gradient approximation within the density functional theory with the PBE (named after Perdew, Burke, and Ernzerhof) exchange-correlation functional \cite{PBE}. The Brillouin zone was sampled by the Gamma point only due to the large size of the supercell. The position of the Nb atoms in the bottom two layers was fixed to their bulk position, while we allowed a vertical geometrical relaxation for the top Nb atoms and the atoms above it (atoms in the overlayer and the magnetic atom). The vertical positions are then determined by averaging within a layer. The magnetic atom does not influence the top Nb and the overlayer relaxation significantly; however, the adatoms' relaxation strongly depends on the material of the overlayer and the magnetic atom. The top Nb layer relaxes towards the vacuum by $2.51\pm 0.02$\% and by $0.62\pm 0.01$\% compared to ideal bulk vertical positions in the case of the Re and Ir overlayers, respectively, while it relaxes towards the bottom Nb layers by $1.02 - 0.95$\% if there is an Ag overlayer. The Re/Ir overlayer relaxes towards the top Nb layer by $13.0$\%/$12.8$\%, respectively, while the Ag overlayer relaxes towards the vacuum by $1.9-2.0$\%. The adatom relaxes towards the surface by $9.7-29.4$\%, depending on the type of the adatom and the material of the overlayer (see the Supplementary Information for the calculated values for each system\cite{SupMat}).

Before performing the embedded cluster calculation, we first performed self-consistent SKKR calculations of the layered systems only, with the appropriate relaxations. The energy integration was performed along a semicircle on the complex energy plane, starting from around 1 Ry below the Fermi energy. The irreducible wedge of the 2D Brillouin zone was sampled by 5151 k-points in all the calculations. Then we embedded the magnetic chains together with neighboring overlayer atoms and vacuum positions within a radius of $1.01 a_{\textrm{Nb}}$ with respect to the magnetic atoms to more accurately account for the changes in the electrostatic potential. Consequently, the embedded cluster contained 41 or 58 ovelayer atoms and 82-136 vacuum positions depending on the crystallographic direction of the 19-atom-long chain. After the embedded cluster was calculated self-consistently with a ferromagnetic $z$ (FM$z$, perpendicular to the surface) alignment of the spins, we determined the magnetic interactions in the extended classical Heisenberg model of the form
\begin{equation}
    \begin{aligned}
    	\Omega\left( \left\{ \vrd{e} \right\} \right)& = \Omega_0 + \sum_{i=1}^N \vrd{e}_i \mxrd{K}_i \vrd{e}_i \\&- \frac{1}{2} \sum_{\substack{
   i,j=1 \\
   i \neq j}}^N J^I_{ij} \vrd{e}_i \vrd{e}_j + \vec{D}_{ij} \left(\vrd{e}_i \times \vrd{e}_j\right)+\vec{e}_i \mxrd{J}_{ij}^S \vec{e}_j,
   \end{aligned}
		\label{eq:heis}
\end{equation}
with the spin-cluster expansion technique\cite{szunyogh2011,deak2011,laszloffy2017}. In \eqref{eq:heis}, $\vrd{e}_i$ is the magnetization direction at site $i$, $J^I_{ij}$ is the isotropic interaction, $\mxrd{K}_i$ and $\mxrd{J}_{ij}^S$ are the one and two-ion anisotropy matrices, respectively, and $\vec{D}_{ij}$ is the Dzyaloshinskii--Moriya vector. Note that close to the ferromagnetic state, the one and two ion anisotropy can be combined as
\begin{equation}
 \mxrd{A}_{i} = \mxrd{K}_{i} - \frac{1}{2} \sum\limits_{j=1}^{N} \mxrd{J}_{ij}^{S}\,.
 \label{eq:anisitefm}
\end{equation}
which can be called the site-resolved anisotropy matrix. 
The normalized eigenvectors of the symmetric matrices $\mxrd{K}_{i}$ or $\mxrd{A}_{i}$, $\vrd{e}^{\,e}_i$, $\vrd{e}^{\,i}_{i}$, and $\vrd{e}^{\,h}_{i}$, correspond to the easy, intermediate, and hard directions, with the respective energy eigenvalues $k^{e}_{i} \leq k^{i}_{i} \leq k^{h}_{i}$. For quantifying the site-specific easy directions together with the magnetic anisotropy energies, we will use the following vector:
\begin{equation}
\vrd{k}_i = \left( k^m_i - k^e_i \right) \vrd{e}^{\,e}_i \,.
\end{equation}

The ground state of the magnetic chains is determined from low-temperature Metropolis Monte Carlo simulations of the spin model, followed by zero-temperature Landau--Lifshitz--Gilbert spin-dynamics simulations started from the final configuration of the Monte Carlo simulations. The details of the simulations are given in Refs.~\cite{laszloffy2017,laszloffy2019}. Most of the systems converged easily, with at least 9 of 10 independent runs resulting in the same energy minimum; however, for a few systems, it was challenging to find the magnetic ground state, and around 10 out of 100 runs resulted in the same energy minimum of the spin system.

Once the magnetic ground state was obtained, we repeated the SCF calculations using that magnetic configuration. Then, as a final step, we calculated the LDOS in the superconducting state around the superconducting gap of Nb: $\varepsilon \in \varepsilon_\textrm{F} \pm 2.04 \textrm{meV}$, sampled by 301 energy points distributed uniformly within this range. We also extended the superconducting state calculations to the ferromagnetic $z$ configuration in all cases. While this may not be the proper ground state, but it may be achieved and controlled by a small magnetic field in an experimental setup.

\section{Results}

\subsection{Overview of magnetic ground states}

In this section, we present the magnetic ground states and superconducting local density of states (LDOS) of transition-metal chains deposited on Ag, Re, and Ir overlayers on the Nb(110) surface. The investigated systems comprise Cr, Mn, Fe, and Co chains oriented along several crystallographic directions, providing a broad set of material combinations for exploring the interplay between magnetism and superconductivity. Our objective is to identify how the magnetic ground state, together with the electronic structure of the substrate, determines the formation of the Shiba bands, superconducting minigaps, and localized zero-energy edge states.

\begin{table*}[t]
\caption{Magnetic ground states of the investigated 19-atom-long chains. Ferromagnetic (FM) systems are grouped in the first column. The alignment of the spins is indicated in the second column by the letters $x$, $y$, or $z$ in the ground-state magnetic configuration. Due to reduced symmetry, the easy axis of two systems lies within the surface plane, denoted by $xy$. Antiferromagnetic (AFM) systems are collected in the third column, together with their corresponding easy-axis directions in the fourth column. A surprisingly large number of systems exhibit spin-spiral ground states; these are listed in the fifth to seventh columns, together with the plane of rotation and the angle between neighboring spins in the middle of the chain as a reference point. We denote the spin orientation by $xyz$ when the magnetic moments possess nonzero components along all three Cartesian directions, i.e., the spins are neither confined to a principal axis nor to a crystallographic plane. Finally, the last column contains two systems with an exotic up-up-down-down magnetic ground state, which is analyzed in detail in Sec.~\ref{sec:exotic}.}
\label{tab1}
\begin{ruledtabular}
\begin{tabular}{ccccccrc}
FM & dir & AFM & dir & Spiral & plane & angle & Other\\
\hline
Cr $x$/Re & $x$&Cr $y$/Ag & $x$&Cr $y$/Re & $yz$ & 14.8&Co $x$/Re \\
Cr $x$/Ir & $x$&Cr $u$/Ag & $x$&Cr $u$/Re & $xyz$ & 170.0&Co $x$/Ag \\
Cr $y$/Ir & $z$&Mn $x$/Ir & $x$&Cr $u$/Ir & $xz$ & 77.0& \\
Mn $y$/Re & $x$&Mn $u$/Ag & $x$&Cr $x$/Ag & $xyz$ & 166.0& \\
Mn $x$/Ag & $x$&Fe $x$/Ir & $y$&Mn $x$/Re & $xz$ & 8.4& \\
Mn $y$/Ag & $x$&Co $y$/Ir & $x$&Mn $u$/Re & $xz$ & 136.7& \\
Fe $x$/Re & $z$&&&Mn $y$/Ir & $yz$ & 73.4& \\
Fe $u$/Re & $xy$&&&Mn $u$/Ir & $xyz$ & 76.9& \\
Fe $y$/Ag & $z$&&&Fe $y$/Re & $yz$ & 20.6& \\
Fe $u$/Ag & $z$&&&Fe $y$/Ir & $yz$ & 92.7& \\
Co $y$/Ag & $y$&&&Fe $u$/Ir & $xyz$ & 92.0& \\
Co $u$/Ag & $xy$&&&Fe $x$/Ag & $xz$ & 38.2& \\
&&&&Co $y$/Re & $yz$ & 137.9& \\
&&&&Co $u$/Re & $xyz$ & 68.3& \\
&&&&Co $x$/Ir & $xy$ & 132.8& \\
&&&&Co $u$/Ir & $xz$ & 114.8& \\
&&&&Co $x$/Re & $xz$ & 7.0& \\
&&&&Co $x$/Ag & $xz$ & 8.0& \\
\end{tabular}
\end{ruledtabular}
\end{table*}

Table~\ref{tab1} summarizes the magnetic ground states obtained for all of the investigated 19-atom-long chains. Even before examining the corresponding superconducting state spectra, the results reveal a remarkable diversity of magnetic configurations. Depending on the chemical composition of the chain, the overlayer material, and the crystallographic orientation, the magnetic ground state may be ferromagnetic, antiferromagnetic, noncollinear spin spiral, or even the exotic up-up-down-down configuration discussed in Sec.~\ref{sec:exotic}. This diversity already indicates that the competition between isotropic exchange interactions, magnetic anisotropy, and Dzyaloshinskii--Moriya interactions is highly sensitive to the local atomic environment, suggesting that the superconducting properties cannot be understood from simplified magnetic models alone.

The most striking observation is the unexpectedly large number of spin-spiral ground states. Nearly half of the investigated systems exhibit some sort of noncollinear magnetic ordering, demonstrating that magnetic frustration is a common feature rather than an exception in these hybrid structures. The spirals span a broad range of rotation angles, from almost ferromagnetic configurations with neighboring spins differing by only a few degrees to nearly antiferromagnetic arrangements approaching $180^\circ$. Moreover, several systems exhibit fully three-dimensional spin textures, with magnetic moments possessing finite components along all three Cartesian directions. Such complex magnetic structures emphasize the importance of fully relativistic first-principles calculations for predicting realistic magnetic ground states.

The distribution of magnetic ground states is remarkably balanced. Cr, Mn, and Fe chains each exhibit five collinear (ferromagnetic or antiferromagnetic) and four noncollinear ground states, demonstrating that no single type of magnetic ordering is intrinsically preferred. Indeed, every magnetic element considered realizes ferromagnetic, antiferromagnetic, and spin-spiral ground states in at least one combination of overlayer material and chain orientation. Co displays the richest magnetic behavior, with spin spirals occurring most frequently and the only two up-up-down-down ground states found in the present study. These results highlight the delicate balance between competing magnetic interactions and their strong dependence on both the overlayer material and the crystallographic orientation.

A complete summary of the magnetic properties and the corresponding LDOS in the superconducting state for all investigated systems is provided in the Supplementary Information\cite{SupMat}. In the following, we focus on several representative and particularly instructive examples that illustrate the different physical mechanisms governing the superconducting state spectra.

\subsection{Local density of states in the superconducting state with ground state magnetic configuration}

Although we studied a rather large number of systems, we found that all the results somewhat trivially fall into one of three categories: (i) trivial minigap, no MZM, (ii) topological minigap with MZM and (iii) filled superconducting gap with no minigap. Strictly speaking, there should be one more category set up, where there is an indication that an even longer chain might support an MZM, however, following this trail would have made the paper extensively long. To keep the discussion reasonably concise, we elected to discuss only one or two representatives from each of the these categories. The complete set of results is summarized in the Supplementary Information\cite{SupMat}.

\begin{figure*}
    \centering
    \includegraphics[width=\linewidth]{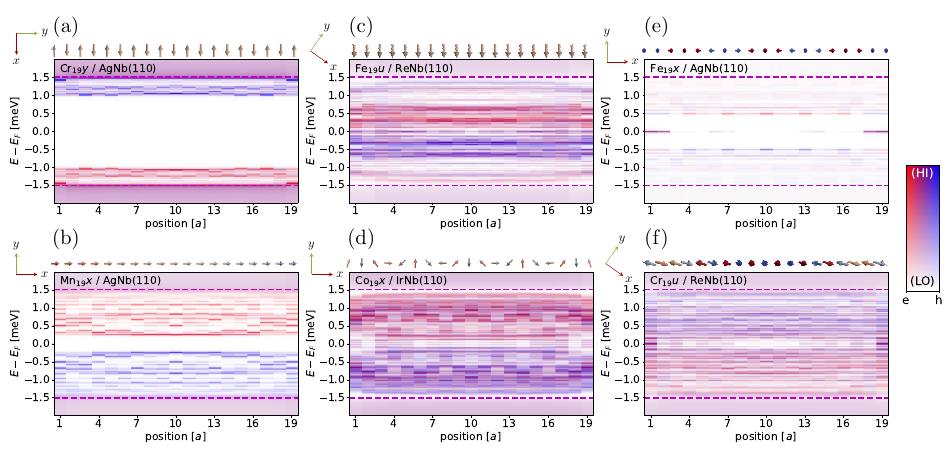}
    \caption{Magnetic ground states and local density of states (LDOS) in the superconducting state for selected magnetic chains deposited on overlayers on the Nb(110) surface and oriented along different crystallographic directions. Red and blue colors represent electron-like and hole-like contributions, respectively, with a continuous transition through purple corresponding to varying electron and hole character. The LDOS is encoded in the color intensities and normalized according to the color bar shown on the right. The two leftmost systems (a-b) exhibit characteristic topologically trivial behavior, with well-defined minigaps and no zero-energy edge states. The systems in the middle column (c-d)  display the superconducting gap that is largely filled with overlapping electron-like and hole-like features. The Fe$_{19}x$ / AgNb(110) (e) chain exhibits characteristic signatures of a topologically nontrivial phase, including a large minigap, strongly localized zero-energy edge states, and mixed electron-hole character on both sides of the Fermi energy. In contrast, the Cr$_{19}u$ / ReNb(110) (f) chain also hosts localized zero-energy edge states, but extended Shiba bands largely fill the superconducting gap.}
    \label{fig1}
\end{figure*}

We begin with a common scenario frequently encountered in antiferromagnetic chains, namely the formation of a topologically trivial minigap (inside the full superconducting gap), where the Shiba bands are well separated, and no localized states appear within the gap.

As an illustrative example, Fig.~\ref{fig1}a shows the Cr$_{19}y$ / AgNb(110) chain, which exhibits an in-plane antiferromagnetic ground state with spins aligned along the $\pm x$ directions. In the corresponding real-space spectrum in the superconducting state, the horizontal axis traces the positions of the magnetic atoms, while the vertical axis represents the energy. The local density of states (LDOS) is encoded in the intensity of each point and normalized according to the legend shown on the right side of the figure. Also, to reduce the number of figures to a manageable amount, the electron and hole contributions are color-coded: red and blue LDOS correspond to electron and hole components, respectively, with a continuous transition between them (purple). This visualization technique allows both electron and hole contributions to be represented simultaneously, without the need to introduce any additional quantities, most importantly the superconducting charge density of states (CDOS in Refs.~\onlinecite{nyari_PRBI,laszloffy_PRBII}).

Returning to the Cr$_{19}y$ / AgNb(110) chain, the electron-like states are located well below the Fermi energy and approach the coherence peak of Nb. Correspondingly, hole-like features appear at positive energies, consistent with the electron–hole symmetry inherent in the solutions of the Bogoliubov–de Gennes equation. In the energy range between $-1$ meV and $1$ meV, no states are present, indicating the existence of an exceptionally large minigap. Since the electron- and hole-like states are well separated at the minigap edge, the band structure can be naturally identified as topologically trivial; consequently, it is not surprising that no zero-energy states emerge within the minigap.

A similar situation is observed for the Mn$_{19}$x / AgNb(110) system in Fig.~\ref{fig1}b; however, in this case, the magnetic ground state is an in-plane ferromagnetic configuration, with the magnetic moments aligned along the direction of the chain. The Shiba bands are significantly broader, likely as a consequence of the stronger hybridization between the magnetic atoms associated with the ferromagnetic ordering. It should also be noted that the electron-like contributions appear above the Fermi level, while the size of the minigap is reduced to approximately 0.24 meV.

While both of the aforementioned systems exhibit particularly large magnetic moments—averaging 4.33 $\mu_\mathrm{B}$ for Cr and 4.59 $\mu_\mathrm{B}$ for Mn—the situation changes significantly when Fe or Co chains are considered. This difference can largely be understood by revisiting how the energy of a single impurity Yu–Shiba–Rusinov (YSR) state evolves as a function of the magnetic moment. For small values of it, YSR states appear close to the coherence peak of Nb. As the magnetic moment increases, the YSR states progressively shift toward the Fermi energy, eventually crossing it, then approaching the opposite coherence peak, and finally merging with it at sufficiently large magnetic moments \cite{yu1965bound,shiba1968classical,rusinov1969superconductivity,Nyari2021}.
In the case of the Fe$_{19}u$ / ReNb(110) chain in Fig.~\ref{fig1}c, the average magnetic moment is 2.80 $\mu_\mathrm{B}$, which corresponds to only about 65\% of the magnetic moment of the previously discussed Cr atoms. In addition, it has already been shown for a single Fe adatom on a Re overlayer on Nb(110) that the YSR states tend to lie deeper within the superconducting gap \cite{laszloffy2023YSR}. The ferromagnetic ground state, together with the small nearest-neighbor distance, promotes strong hybridization, leading to broad YSR bands. Consequently, it is not surprising that these extended YSR states fill the superconducting gap.

The Co${19}x$ / IrNb(110) chain in Fig.~\ref{fig1}d exhibits a similar behavior in the sense that its average magnetic moment is 2.18 $\mu\mathrm{B}$, and the YSR states of a single adatom are distributed more or less uniformly within the superconducting gap. The magnetic ground state, however, is more complex: it forms a spin spiral with a periodicity of eight atoms. At the center of the chain, the nearest-neighbor (NN) isotropic exchange interaction is $-1.97$ meV, while the next-nearest-neighbor (NNN) isotropic interaction is $-0.85$ meV. Both interactions correspond to antiferromagnetic coupling, leading to magnetic frustration. The magnitude of the NN Dzyaloshinskii–Moriya (DM) vector is 2.26 meV, which is even larger than the isotropic exchange interactions. However, its direction is parallel to the easy axis, which lies along the $y$ direction, with a medium-easy anisotropy energy difference of $k_{10}^m - k_{10}^e = -0.62$ meV. Consequently, the plane in which the spins rotate is tilted away from the $x$–$y$ plane, allowing the DM interaction to contribute to the resulting eight-atom periodicity of the spin spiral.

The remaining question for these two systems (Fe$_{19}u$ / ReNb(110) and Co${19}x$ / IrNb(110) chains) is why a minigap does not emerge together with the appearance of zero-energy states, even though spin–orbit coupling (SOC), in principle, could induce such a feature. Several factors may contribute to this behavior. The magnetic orientation may not be favorable for effective cooperation with SOC, the chain length may not be long enough, or some of the bands may approach the Fermi level without actually crossing it within our multiband description.

Among the investigated systems, Fe$_{19}x$/AgNb(110), shown in Fig.~\ref{fig1}e, is perhaps the most compelling example, as it exhibits all the characteristic features of a topological MZM. In the superconducting state LDOS, a particularly large minigap of 0.5 meV is present around the Fermi energy, accompanied by zero-energy states that are almost perfectly localized at the edges within approximately two atomic sites. Furthermore, signatures of band inversion can be identified on the central atoms in the state at 0.5 meV: (i) both electron-like and hole-like features are observed on either side of the superconducting gap, (ii) the interior of the chain is dominated by electron-like contributions, with a transition to hole-like character toward the edges.

It is somewhat surprising that the magnetic ground state is a spin spiral. The isotropic magnetic interactions in the middle of the chain are characterized by $J_{10,11}^{\mathrm{I}} = 2.37\mathrm{meV}$ and $J_{10,12}^{\mathrm{I}} = -1.37\mathrm{meV}$. The easy axis is along the $z$ direction, with a medium-easy anisotropy energy difference of $1.68,\mathrm{meV}$, while the amplitude of the nearest-neighbor Dzyaloshinskii–Moriya (DM) interaction is relatively small; the amplitude of the DM vector is $0.3\mathrm{meV}$. The competition between nearest-neighbor ferromagnetic and next-nearest-neighbor antiferromagnetic interactions leads to magnetic frustration, which stabilizes a spin spiral ground state with a periodicity of six atoms. The plane of rotation of the spins, as well as the chirality of the spiral, is determined by the combined effects of magnetic anisotropy and the DM interaction.

The final system shown in Fig.~\ref{fig1} is the Cr$_{19}u$ / ReNb(110) chain, which exhibits an almost antiferromagnetic ground state with a modulation wavelength exceeding the length of the chain. This behavior can be readily understood by considering the hierarchy of the spin model parameters, as detailed in the Supplementary Material\cite{SupMat}. Despite the predominantly antiferromagnetic character, zero-energy states can be observed, localized at the edges of the chain. At the same time, the superconducting gap appears to be largely filled, with only a small residual gap that is significantly less than 0.1 meV. The broadening of the YSR bands can be attributed to the strong hybridization between the magnetic atoms, sitting very close to each other along the [$\overline{1}11$] crystallographic direction, which are in close proximity to one another. This system can therefore be classified as one exhibiting zero-energy states in the absence of a well-defined minigap. Such behavior may arise within a multiband framework, where multiple bands are present: some may host a minigap and corresponding zero-energy states \cite{nyari2025}, while others remain topologically trivial and do not exhibit such features.

\subsection{Chains constrained into a ferromagnetic order}

\begin{figure*}
    \centering
    \includegraphics[width=\linewidth]{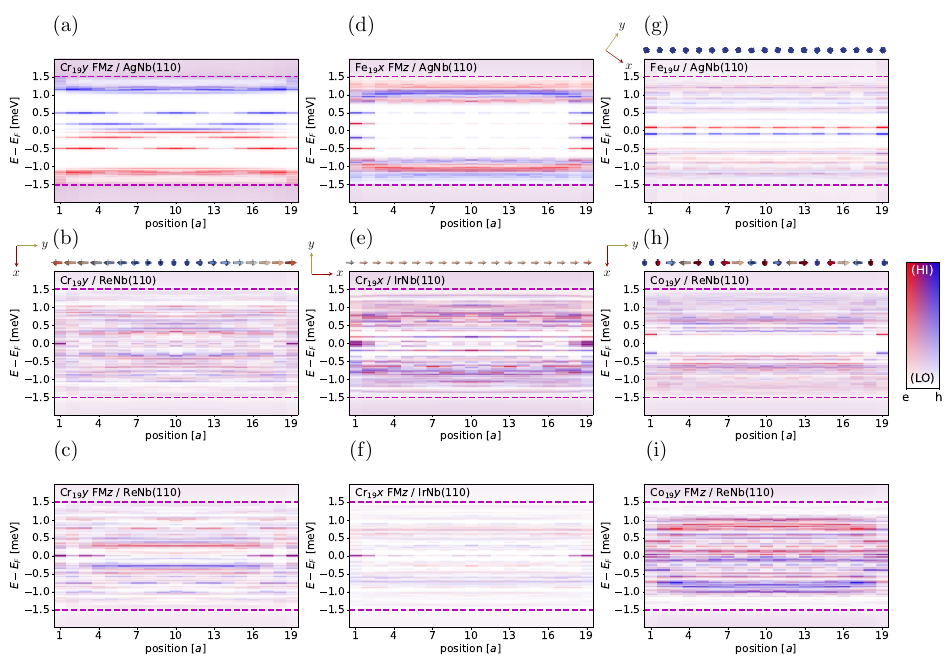}
    \caption{Effect of the magnetic state on the 
    superconducting local density of states (LDOS) for selected magnetic chains 
    on overlayers on the Nb(110) surface. Whenever the magnetic ground-state configuration is shown above the LDOS, the spectrum corresponds to the calculated magnetic ground state. In contrast, panels labeled by ``FM$z$'' (e.g., Cr$_{19}y$ FM$z$ / ReNb(110)) correspond to calculations in which the magnetic moments were constrained to a ferromagnetic configuration aligned along the $z$ direction. 
    Comparing the ground-state and FM$z$ configurations highlights how magnetic ordering reshapes Shiba bands, modulates superconducting minigaps, and controls the appearance of zero-energy edge states.}
    \label{fig2}
\end{figure*}

In this section, we investigate how the Shiba band structure evolves when the system is forced into a state with strict ferromagnetic ordering. This can be achieved in principle by applying an appropriate magnetic field. If this magnetic field is smaller than the superconducting critical field, such manipulation -- in principle -- can be performed in experiments. An estimate of the magnetic field strength necessary to make the ferromagnetic $z$ configuration more favorable is given in the Supplementary Information for each studied system\cite{SupMat}. 
To estimate the magnetic field where the ferromagnetic $z$ configuration becomes energetically favorable -- denoted by $B_\mathrm{rot}$ -- a Zeeman term was added to the spin Hamiltonian (\ref{eq:heis}). In most cases, when the ground state is a spin spiral or the easy direction is in the plane. (not to mention the antiferromagnetic ground state), this field exceeds 10T. There are, nevertheless, notable exceptions, where $B_\mathrm{rot}$ is comparable to the superconducting critical field, for example, in the Fe$_{19}x$ / AgNb(110) system, $B_\mathrm{rot}=1.29 \mathrm{T}$. Anyway, by comparing the superconducting state LDOS of the appropriate ground state and the ferromagnetic $z$ configuration, we can identify some interesting features. It is necessary to mention that the external field was not included in the ab-initio calculations, only the moment directions were forced into a ferromagnetic alignment in the perpendicular direction.

If the ground state is antiferromagnetic, as in the Cr$_{19}y$ / AgNb(110) system (see Fig. \ref{fig1}a), transforming the magnetic configuration into a ferromagnetic state enhances the hybridization between the magnetic atoms, which in turn can lead to a significant broadening of the Shiba bands. Indeed, as shown in Fig.~\ref{fig2}a, bound states with a well-defined number of maxima emerge within the interior of the superconducting gap, at energies where no states were present in the (antiferromagnetic) ground state. 

Probably the most interesting question is how the LDOS evolves for the Fe$_{19}x$ / AgNb(110) chain, where previously a topological edge state was predicted in the ground state (Fig. \ref{fig1}e). If the ferromagnetic state is enforced (see Fig.~\ref{fig2}d), a much larger minigap of 0.8 meV can be observed, with two pairs of YSR states localised at the edges. The one that is closer to the Fermi energy has a large electron component above the Fermi level, while the other one at 0.5 meV has a hole-like character. To clarify the topological nature of the extremely large minigap, we performed calculations without spin-orbit coupling \cite{SOCscaleHubert} as well. Unfortunately, the calculations without SOC display very similar features; consequently, the edge states appear to have no Shiba band inversion and are topologically trivial. The corresponding edge states are just edge-localized YSR states. When the chain is in a different crystallographic direction, like the Fe$_{19}u$ / AgNb(110) chain, the large minigap of 0.5 meV remains, but only one pair of high-intensity state at 0.095 meV remains in the minigap with 8 maxima, and with a definite electron or hole contribution. A close inspection suggests the existence of a faint sign of topological LDOS around 0.31 meV in the interior of the chain. This system is a prime example of the case where an even longer chain may support a topological edge state. 

The Cr${19}y$ / ReNb(110) chain shows a transition from a filled gap in the ground state to a topologically non-trivial state in the ferromagnetic configuration (Fig. \ref{fig2}b-c). The magnetic ground state in this case is a spin spiral characterized by a very small angle between neighboring spins, and it belongs to the category where the superconducting gap is entirely filled, with both electron-like and hole-like components observable on either side of the Fermi level. Despite the gap being filled, a clear signature of zero-energy states may be observed, which are not strictly localised to the edge atoms. As noted previously, this behavior may be associated with a topological band coexisting with other bands that remain topologically trivial. However, the absence of a well-defined minigap makes this system impractical for applications. Enforcing a ferromagnetic configuration -- keeping in mind that the ground state is already close to a ferromagnetic alignment -- leads to a reduced spectral density in the vicinity of the Fermi level, while the zero-energy edge states remain present and even become more localised. This suggests that, in such cases, the ferromagnetic state may be more favorable for the formation of well-defined zero-energy edge states than spin spirals with small spiraling angles. 

In the $x$ direction the same chain (Cr$_{19}x$ / IrNb(110), Fig. \ref{fig1}e-f) demonstrates a clear anisotropy effect. While the in-plane ferromagnetic ground state magnetization does not support the formation of a minigap, rotating the spins toward the $z$ direction leads to the emergence of a minigap of 0.16 meV, along with localized edge states that exhibit long decay tails extending toward the center of the chain. This effect has already been identified in the case of a Mn chain on superconducting Ta.\cite{nyari2025} Note that edge states close to the Fermi energy can also be identified with the in-plane magnetization; however, there are additional extended YSR states in the vicinity of the Fermi energy.

The Co$_{19}y$ / ReNb(110) chain exhibits a spin spiral ground state characterized by an angle of $137.9^\circ$ between neighboring spins in the middle of the chain. This corresponds to a spin spiral with a wavelength of approximately 2.61 atoms, which means the spins rotate around the $x$ axis roughly three times over a distance of eight atoms. In this spin spiral configuration, the superconducting state spectrum shown on Fig. \ref{fig2}h displays a relatively broad minigap of 0.26 meV, while both electron-like and hole-like contributions are present on either side of the Fermi energy. However, there are no in-gap states. For the ferromagnetic $z$-aligned configuration (Fig. \ref{fig1}i), the minigap closes. Nevertheless, both electron-like and hole-like contributions remain observable at comparable energies on either side of the Fermi energy. This behavior likely indicates the presence of a broad band overlapping the Fermi level, or alternatively, the coexistence of multiple bands contributing to the overall band structure. The remaining question is how the spin-spiral configuration can open a gap in this system without generating zero-energy edge states. One possible explanation is that the spectral function crosses the Fermi level twice, resulting in a vanishing winding number \cite{nyari2025}. This interpretation is appealing because it naturally accounts for the opening of the minigap in the spin spiral state while simultaneously explaining the absence of zero-energy edge states.

\subsection{Exotic magnetic ground states and their impact on superconducting LDOS}
\label{sec:exotic}

\begin{figure*}
    \centering
    \includegraphics[width=0.7\linewidth]{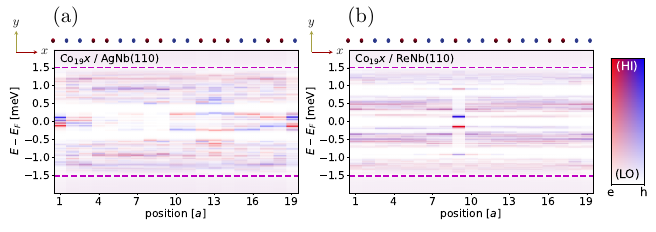}
    \caption{Magnetic ground states and superconducting local density of states (LDOS) of Co$_{19}x$ chains on Ag and Re overlayers on the Nb(110) surface. Both systems exhibit an exotic up-up-down-down magnetic ground state with a localized magnetic domain wall. The corresponding superconducting LDOS is shown below each magnetic configuration.}
    \label{fig3}
\end{figure*}
During our spin-dynamics calculations, we discovered a collinear, up-up-down-down (uudd) magnetic ground state can be stabilized within a second-order Heisenberg model, without including higher-order spin interactions. These systems are unique because of the consequences when a chain length is incommensurate with this exotic magnetic ordering. Among all investigated systems, this ground state was observed in two cases, both involving Co chains oriented along the side-diagonal $x$ direction, with either Ag or Re as the overlayer. A summary of the corresponding spin model parameters is provided in the Supplementary Information\cite{SupMat}. In the following, the two systems are analyzed separately.

For the Co${19}x$ / AgNb(110) chain, the hierarchy of the Heisenberg model parameters is as follows. The easy axis points perpendicular to the surface, with a magnetic anisotropy energy of 3.77 meV per atom. In the middle of the chain, the nearest-neighbor (NN) isotropic interaction is $J_{10,11}^{I} = 0.59,\mathrm{meV}$, corresponding to ferromagnetic coupling, while the next-nearest-neighbor (NNN) isotropic interaction is $J_{10,12}^{I} = -0.40,\mathrm{meV}$, corresponding to antiferromagnetic coupling. The magnitude of the NN Dzyaloshinskii–Moriya (DM) interaction is 0.52 meV; however, its effect on this chain is negligible. Considering only the NN and NNN isotropic interactions, magnetic frustration favors a spiral ground state. However, the magnetic anisotropy energy is approximately seven times larger than the isotropic exchange interactions, forcing the spins to align predominantly along the $\pm z$ direction. As a result, the strong anisotropy stabilizes the collinear up-up-down-down configuration.

This behavior can also be understood from an energetic perspective. In a collinear ferromagnetic configuration, the contribution of the isotropic interactions to the energy per atom is simply given by the sum of the interaction energies, taking into account that each bond is shared equally between neighboring spins. This yields
$E^I_{\mathrm{FM}} =- J_{10,11}^I-J_{10,12}^I=-0.19 \mathrm{meV}$
per atom for the ferromagnetic state. In contrast, for the up-up-down-down configuration, the corresponding isotropic interaction energy per atom becomes
$E^I_{\mathrm{uudd}}=J_{10,12}^I = -0.40\mathrm{meV}$
which is lower than $E_{\mathrm{FM}}^{I}$. Consequently, the up-up-down-down state is energetically favored.

For homogeneous infinite chains, one would therefore expect a perfect uudd magnetic ground state. However, the finite chain length of 19 atoms is not divisible by four and is thus incommensurate with the periodicity of the magnetic ordering. Furthermore, in finite chains the spin model parameters are not perfectly homogeneous. In the present case, the antiferromagnetic alignment of the two leftmost and two rightmost atoms is energetically more favorable than the uudd configuration found in the interior of the chain. Consequently, three consecutive spins become aligned parallel to one another. Since this local uuu configuration cannot be accommodated symmetrically at the center of the chain, it is shifted away from the middle and centered at the 13th atom.
To verify this argument, we also performed calculations for chains consisting of 18 Co atoms, for which no domain wall appeared in the magnetic ground-state configuration.

On the Re overlayer, the magnetic ground state is qualitatively similar, and, based on the spin-model parameters, for an infinite chain a uudd configuration is again expected. For the finite chain, however, the inhomogeneity of the spin model parameters favors ferromagnetic alignment at the chain edges. Additionally, the udu domain wall is energetically preferable to the uuu domain wall observed in the previous case. This udu configuration also cannot be positioned symmetrically at the center of the chain if other parts are uudd aligned, and is therefore shifted by one atomic site toward the left. It should be noted that, in both systems, applying mirror symmetry with respect to a plane perpendicular to the chain leaves the energy of the spin configuration unchanged. Consequently, the magnetic ground states occur in degenerate pairs, providing an interesting example of spontaneous symmetry breaking in a magnetic system.

The superconducting LDOS reflects this broken symmetry through the appearance of localized YSR states in the vicinity of the domain wall. For the Ag overlayer, a minigap around $E-E_\mathrm{F} = 0.4 \mathrm{meV}$ can be identified in Fig. \ref{fig3}a. In the case of the Re overlayer, apart from the domain-wall states, narrow and energetically well-separated YSR bands can be observed (Fig.~\ref{fig3}b).

\subsection{Effect of spatial relaxation}

\begin{figure*}
\centering
\includegraphics[width=\linewidth]{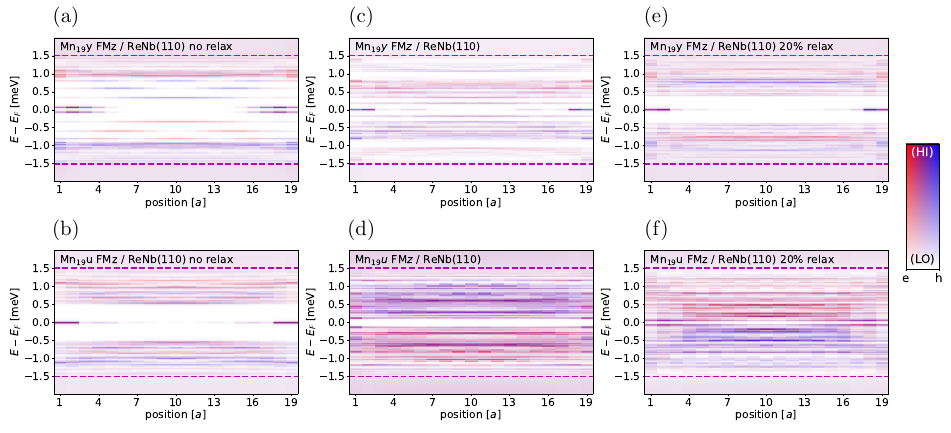}
\caption{Effect of spatial relaxation on the superconducting local density of states (LDOS) of Mn chains on the Re overlayer on the Nb(110) surface. For all calculations, the magnetic moments were constrained to a ferromagnetic configuration aligned along the $z$ direction (FM$z$). The upper and lower rows correspond to chains oriented along the $y$ and $u$ crystallographic directions, respectively. The left column shows the LDOS of the unrelaxed geometry with atoms fixed at their bulk position, the middle column corresponds to the geometry obtained from VASP calculations, while the right column shows the LDOS with an artificial 20\% inward relaxation of both the Re overlayer and the Mn atoms.}
\label{fig4}
\end{figure*}

As we discussed earlier, both the ground state and the LDOS in the superconducting state is very sensitive to hybridization effects. Therefore, it is expected that any changes in the spatial arrangement of the overlayers and the chain atoms have a profound effect on the LDOS. We demonstrate this on the example of Mn chains deposited on a Re overlayer on the Nb(110) surface, where the chains are oriented either along the $y$ or the $u$ direction. In this analysis, the magnetization direction was kept fixed in the ferromagnetic $z$ configuration to isolate the effects of geometrical relaxation. 

The results of our calculations are shown on Fig.~\ref{fig4}, where the middle column corresponds to the relaxed geometry obtained from VASP calculations in which the Re overlayer relaxes by 13\% toward the bulk, while the Mn atoms relax by 19\% toward the surface. This corresponds to the ground state. In the right column, an artificial relaxation was imposed, where both the Re overlayer and the Mn atoms shifted by 20\% toward the bulk. In the left column, no geometrical relaxation was applied, and all atoms remain at their ideal bulk positions. This arrangement of the columns represents an increasing amount of relaxation toward the bulk, from 0 to 20\% from left to right.
It is interesting to see that the relaxation has an entirely opposite effect on these two chains.
Interestingly, for chains oriented along the $y$ direction, localized edge states are observed in all three cases. In the unrelaxed case they are not entirely at zero energy, and various states can be observed around. Relaxation creates a small, but definite minigap with signs of topological band inversion that appears on the middle atoms. This is probably because hybridization pushed the states that were around zero -- when unrelaxed -- to be dissolved in the minigap. Correspondingly, the edge states, which than appeared just a little apart, are now exactly at zero energy. The localization lengths of the zero-energy edge states and the corresponding topological minigaps are summarized in Table~\ref{tab2}. Further increasing the relaxation expands the minigap to about 0.43 meV and strengthens its topological properties, which is manifested in an even more pronounced edge state. 

Quite the opposite can be observed in the chains oriented along the $u$ direction on the Re overlayer.
In the case of zero relaxation, we obtained a remarkably large topological minigap of 0.54 meV supporting a strong zero-energy edge state localized to approximately two atomic sites. Increasing the relaxation to the ground state, one can see that the minigap has shrunk, its topological properties are gone, and concomitantly the edge state at zero energy also disappeared. The enhanced hybridization means that the previously observed band inversion is now destroyed. By further increasing the relaxation towards the bulk to 20\%, the increasing hybridization further decreases the minigap. 

\begin{table}
\caption{Localization length of the zero-energy edge states on Fig.~\ref{fig4}a,b,c,e and the size of the corresponding minigap for Mn chains on the Re overlayer for different crystallographic chain directions and geometrical relaxations.}
\label{tab2}
\begin{ruledtabular}
\begin{tabular}{cccc}
Chain direction & Relaxation & Localization [$a$] & $\Delta_{\mathrm{mini}}$ [meV] \\
\hline
$y$ & 19\% (VASP)  & 2 & 0.18 \\
$y$ & 20\%  & 2 & 0.43 \\
$y$ & 0\%   & 4 & 0.33 \\
$u$ & 0\%   & 2 & 0.54 \\
\end{tabular}
\end{ruledtabular}
\end{table}

In conclusion, even small uncertainties in the geometrical relaxation can lead to substantial changes in the superconducting band structure, including the formation or suppression of minigaps, their topological properties, and the appearance or disappearance of zero-energy edge states. This result may explain the volatility in experimental efforts as well. Slightly different growth conditions may result in slightly different relaxations, which in turn makes the observation of an MZM even more difficult. 

\section{Conclusion}

In this work, we systematically investigated the magnetic ground states and superconducting local density of states (LDOS) of transition-metal chains deposited on superconducting Nb(110) surfaces with Ag, Re, and Ir overlayers. Using a multiband description of Yu–Shiba–Rusinov states combined with atomistic spin-model calculations, we analyzed the interplay between magnetic ordering, spin–orbit coupling, hybridization, and geometrical relaxation in determining the superconducting band structure and the emergence of zero-energy edge states.

We demonstrated that the investigated systems exhibit a remarkably wide variety of magnetic ground states, including ferromagnetic, antiferromagnetic, spin-spiral, and exotic up-up-down-down configurations. The resulting superconducting spectra range from topologically trivial minigaps without in-gap states to systems displaying pronounced zero-energy edge states. Antiferromagnetic chains such as Cr$_{19}y$ / AgNb(110) were shown to have large, topologically trivial minigaps. In contrast, systems with reduced magnetic moments, such as Fe and Co chains, tend to produce broad Shiba bands that partially or completely fill the superconducting gap due to stronger hybridization.

Particular attention was devoted to systems exhibiting signatures associated with topological superconductivity. The Fe$_{19}x$ / AgNb(110) chain was identified as a particularly promising candidate, displaying a sizable minigap together with strongly localized zero-energy edge states and signatures of band inversion. In contrast, several other systems exhibited zero-energy edge states without a well-defined minigap, which, within a multiband framework, can be interpreted as systems where topologically trivial and nontrivial bands coexist. Our results indicate that the opening of a minigap is highly sensitive to the magnetic configuration and may depend on subtle details such as the relative orientation between spin texture, the spin–orbit coupling or the number of Fermi-level crossings contributing to the winding number. We found no systems where a localized edge state appears in chains along two different crystallographic directions. While this is quite easily understood from the sensitivity of an MZM to the magnetic state, it makes the construction of a "T" or "X" junction for qubits even more challenging.

We further investigated the effect of manipulating the magnetic configuration through artificial ferromagnetic states. In most cases, transforming antiferromagnetic or spin-spiral ground states into ferromagnetic configurations substantially modified the Shiba band structure, altering the minigap and the localization properties of edge states. In particular, we identified examples where spin spirals support a minigap that disappears in the ferromagnetic configuration, as well as systems where ferromagnetic alignment enhances the visibility of zero-energy edge states. These findings show that the superconducting LDOS is highly tunable by manipulating the spin texture.

An additional, interesting result of this work is the stabilization of exotic up-up-down-down magnetic ground states within a purely second-order Heisenberg model, without invoking higher-order spin interactions. We showed that the competition between nearest- and next-nearest-neighbor exchange interactions, combined with strong magnetic anisotropy, naturally stabilizes such collinear frustrated states. For finite chains, the incommensurability between the chain length and the magnetic periodicity leads to spontaneous symmetry breaking and the formation of localized magnetic domain walls, which are directly reflected in the superconducting LDOS through the appearance of localized YSR states.

Finally, we demonstrated that geometrical relaxation has a strong impact on the superconducting band structure. Even relatively small variations in atomic positions can strongly modify the size of the minigap, the localization length of edge states, and the existence of zero-energy modes. Depending on the relaxation geometry, the same system may exhibit either a large minigap with localized edge states or a nearly gapless spectrum without topological signatures. This sensitivity highlights the importance of accurately determining the atomic structure in realistic modeling of magnetic-superconducting hybrid systems. 

Overall, our results reveal the complex and delicate interplay between magnetic frustration, spin–orbit coupling, superconductivity, and atomic geometry in determining the electronic structure of magnetic chains on superconducting surfaces. The broad range of behaviors identified in this work demonstrates both the challenges and the opportunities associated with engineering topological superconductivity in realistic multiband systems. We hope that these insights will facilitate the experimental realization and application of magnetic-chain systems for topological superconductivity. \\

\section*{Acknowledgements}

The authors gratefully acknowledge Levente Rózsa and Gábor Csire for stimulating discussions. We are also indebted to Ádám Gali and his research team for technical guidance concerning the execution of the VASP calculations. This work was supported by the National Research, Development and Innovation Office (NKFIH, Hungary) under Projects FK142601 and K142652. The authors acknowledge Digital Government Development and Project Management Ltd. for providing access to the Komondor high-performance computing (HPC) facility in Hungary.

\bibliography{main}

\end{document}